\documentstyle[prl,aps,epsf]{revtex}

\def\mj2{{M_{ij}^{(2)}}}
\begin{document} 
\draft 
\twocolumn[\hsize\textwidth\columnwidth\hsize\csname %
@twocolumnfalse\endcsname
\phantom{xx}
\vskip -1cm
\hfill{\small {\it Preprint} NCU/CCS-1998-0920; NSC-CTS-981002} 
\medskip

\title
{Origin of the Native Driving Force for Protein Folding}

\author{Zi-Hao Wang$^1$ and H.C. Lee$^{1,2}$}

\address{
$^{1}$Department of Physics and Center for Complex Systems, 
  National Central University, Chung-li, Taiwan 320, ROC\\
and\\ 
$^{2}$National Center for Theoretical Sciences, P.O. Box 2-131, 
Hsinchu, Taiwan 300, ROC} 
\date{Received December 2, 1998; revision received July 28, 1999}
\maketitle

\begin{abstract} 
We derive an expression with four adjustable parameters that
reproduces well the $20\times 20$ Miyazawa-Jernigan potential matrix
extracted from known protein structures.  The numerical values of the
parameters can be approximately computed from the surface tension of
water, water-screened dipole interactions between residues and water
and among residues, and average exposures of residues in folded
proteins.
\end{abstract} 
]

Protein structure and design is a very important topic in life science
where physics and mathematics are indispensable to its understanding
\cite{Wolynes}. Recently Li {\it et al.} \cite{Li97} pointed out some
highly interesting and unexpected properties of Miyazawa and
Jernigan's $20\times 20$ potential matrix ($M$) for protein structure
\cite{Miyazawa85}\cite{Miyazawa96}. This matrix, whose elements are
statistically deduced pair-wise interaction potential energies
among the twenty types of amino acids in proteins 
of known structure, has been
widely applied to protein design and folding simulations
\cite{Jernigan}\cite{Shakhnovich}\cite{Pande}. Li {\it et al.} noticed
that $M$ has a highly accurate leading principal-component
representation: variations of the elements of $M$ from their mean can
be expressed in terms of only the two leading eigenvalues of $M$ and
the eigenvector $\vec q$ of the leading eigenvalue such that 
\begin{equation}
M_{ij} \cong   c_2 q_i q_j + c_1 (q_i + q_j) + c_0,
\label{li_eq}
\end{equation}
where $i$ and $j$ label the 20 amino acids, 
and $c_0 = -1.38$, $c_1= 5.03$ and $c_2 = -7.40$, in units of $RT$, 
the gas constant times (room) temperature.  

Two features of the right-hand-side of Eq.~(\ref{li_eq}) stand out: 1)
Not all residue-dependent terms are genuine two-body interactions; the
$c_1$ terms represent one-body, mean-field potential energies.  2)
Both the two-body $c_2$ terms and the one-body $c_1$ terms depend on
the same set of $q$'s.  Numerically, because the magnitudes of the
$q$'s are small, the $c_1$ terms dominate over the $c_2$ term.  This
is consistent with the widely held notion that the earliest and
fastest part of a protein folding process is by and large controlled
by the hydrophobicity \cite{Tanford} of the residues.  Tables I and II
show that indeed $q$ is moderately correlated with the
hydrophobicities ($\Delta G$) \cite{Roseman}.  The product, pairwise
form of the two-body terms reminds one of dipole-dipole interaction,
and this in turn would imply a connection between the one-body terms
and the dipole moments of the residues.  Tables I and II also show a
noticeable correlation between $q$ and the dipole moments ($Q$) of the
side-chains of the residues \cite{Chipot}.  In the rest of the paper
we will derive an expression for the MJ matrix in terms of an average
``bare'' residual solvation energy (for a hypothetical residue with
vanishing dipole), interactions between the dipole moments of the
residues and water molecules, and the degree of exposure to water
(expressed as its complement, the burial factor) of a residue in a
folded protein.  We show that except for the burial factor of the
residues the other three adjustable parameters appearing in the
expression all have clear physical meanings with numerical values that
can be computed approximately.  The average burial factors for
hydrophobic and hydrophilic residues that emerge from our analysis of
the MJ matrix are 0.8 and 0.2, respectively (they are related and
should approximately sum to 1).  In this paper energy will be given in
units of $RT = 0.60~ {\rm kcal/mol} = 4.2 \times 10^{-21}~J$ and
dipole moments will be given in Debyes ($D$).

\medskip 
\noindent{\bf Dipole-dipole interaction}.  
The interaction in vacuum 
between two electric dipoles $\vec Q_i$ and $\vec Q_j$ 
separated by $\vec R_{ij}= \hat n R_{ij}$ is
$V_{ij}  = (\vec Q_i \cdot \vec Q_j 
- 3(\hat n\cdot \vec Q_i) (\hat n\cdot \vec Q_j)) 
/(4 \pi \epsilon_0 R_{ij}^3)$.
If the carriers of the dipoles are relatively unconstrained we expect 
attraction and 
$- |\mu_r| |Q_i| |Q_j| \le V_{ij} \le 0$, 
where $|\mu_r| = D^2/ 2 \pi \epsilon_0 R_{ij}^3$.  
In what follows, $Q_i$, $i=1,\cdots,20$ is the dipole moment of the
$i^{th}$ side-chain, and $Q_w$ is the dipole moment of a water
molecule.  For residue-residue interaction, taking the
inter-side-chain distance to be $R_{ij} \cong R_0 \cong 6.5~\AA$
\cite{Miyazawa85}, and recalling that an electron-positron pair
separated by one $\AA$ is equal to 4.8 $D$, we have $|\mu_r| \approx
0.172$ $(RT)$, which may be viewed as a maximum value for
the coupling since in a real setting it is
expected to be weakened owing to the presence of water molecules.

\medskip
\noindent{\bf One-body terms}.  
Let $E_0$ be the average bare surface-dependent 
solvation energy of a residue in water
when the residue-water dipole interaction is not taken into account;
$N_w$ be the average number of water molecules in contact with a
residue; $\mu_w$ be the average effective dipole-dipole coupling
between the $i^{th}$ residue and a water molecule.  Then, with
residue-water interaction energy included and possible dependence of
$E_0$, $\mu_w$ and $N_w$ on $i$ ignored, the residue-water interaction
energy is
$E_i = \mu_w Q_i Q_w N_w + E_0 \equiv \mu_w Q^*_i Q_w N_w$, 
where for convenience we write $Q^*_i \equiv Q_i +Q_0$ and $Q_0 =
E_0/(\mu_w Q_w N_w)$.  A hydrophobic (hydrophilic) residue would have
$E_i > 0$ ($E_i < 0$).  If $N_i$ is the number of the type $i^{th}$
residues in a peptide, then the energy of an unfolded peptide in water
is $U = \sum_i N_i E_i$.  Suppose that after folding $\Delta N_i$
fewer $i^{th}$ residues are exposed to water.  Then the binding energy
of the folded relative to the unfolded state is $\Delta U = - \sum_i
\Delta N_i E_i$.  The negative sign means that in folding, the peptide
will maximize (minimize) those $\Delta N_i$ whose $E_i$ are the most
positive (negative), subject to the constraint of polymeric nature of
the peptide.

\vspace{8pt}
\begin{figure}[hbt]
\epsfxsize=7.6cm  
\epsfysize=9cm 
\epsfbox{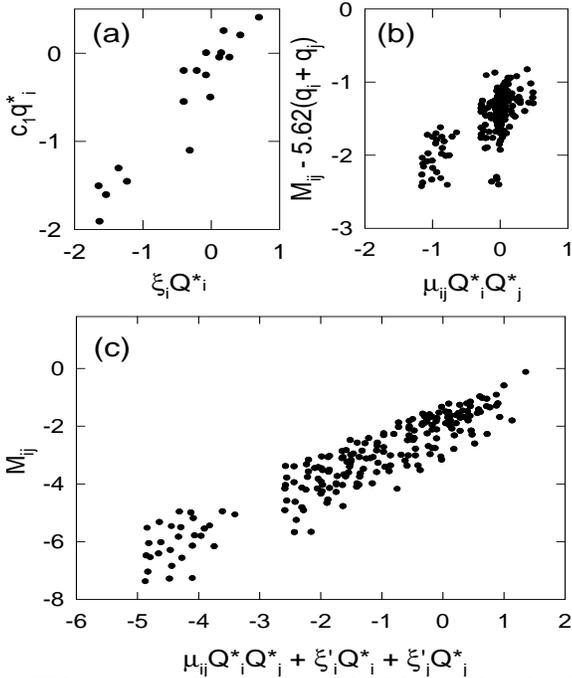}  
\noindent
\caption{\label{figure1}\small 
(a) $\xi_i Q^*_i$ {\it vs.} $c_1 q^*_i$.
(b) The residue dipole-dipole interaction {\it vs.} the two-body term 
in the MJ matrix. 
(c) The right-hand side of Eqs.(\ref{final_eq}) {\it vs.} the complete 
MJ matrix.} 
\end{figure}

\medskip
\noindent{\bf Relation between $q$ and $Q$}. 
Equating $\Delta U$ with the  binding energy obtained from 
Eq.~(\ref{li_eq}) by summing the one-body terms over all pairs we have  
\begin{equation}
\Delta U \approx c_1 N_c \sum_i N_i q^*_i 
= - \mu_w  Q_w N_w \sum_i Q^*_i \Delta N_i
\label{energy2_eq} 
\end{equation}
where $q^*_i \equiv q_i-q_0$, $q_0$ is a constant and $N_c$ 
is the average number of contacts a residue has in a folded state.  
Matching the  $i$-dependent terms we have 
\begin{equation}
c_1 q^*_i  \propto \xi_i Q^*_i,\quad 
\xi_i = -\mu_w (\Delta N_i/ N_i) (N_w Q_w/N_c).
\label{xi_eq}
\end{equation}
Because in a folded protein proportionally more hydrophobic ($h$)
residues than polar ($p$) residues will be hidden from water, one
expects $\Delta N_i/ N_i$, hence $\xi_i$, to have a strong residual
dependence.  To minimize the number of parameters we allow $\xi_i$ to
have only two values: $\xi_h$ and $\xi_p$, and have them determined by
separate linear fits to $q$'s belonging to hydrophobic and hydrophilic
residues, respectively.  Excluded in the fits are residues whose
hydrophobicities are ambivalent \cite{liwx} - Tyr, Ala, Gly, Thr, Ser
and Pro.  Demanding that the two fits have the same intercepts we
obtain
\begin{equation}
q_0 = -0.055,\ Q_0 = - 2.9;\ \xi_h = 0.56,\ \xi_p = 0.14
\label{xi_ex_eq}
\end{equation}
The linear correlation between $q$ and $\xi Q^*$ over
the {\it complete} set of 20 residues - 
following \cite{Li97} and \cite{Keskin}, the first eight
amino acids in Table I are taken to be hydrophobic - 
is 0.949, which is dramatically better than the correlation 
between $q$ and $Q$, see Fig.~1(a) and Table II.

\medskip
\noindent{\bf The burial factor}. 
Since on average the numbers of hydrophobic and polar residues in a 
protein are approximately equal and about half of all residues are 
buried in the core, we have  
$N_h \approx N_p$, $\Delta (N_h + N_p)/(N_h + N_p) \approx 1/2$ 
and hence  $\Delta N_p/N_p \approx 1- \Delta N_h/N_h$.  
From the ratios of the two $\xi$'s we thus deduce the burial factors 
for hydrophobic and polar residues, respectively, to be
\begin{eqnarray}
\Delta N_h/ N_h \approx 0.80,\qquad
\Delta N_p/ N_p \approx 0.20. 
\label{hidden_eq}
\end{eqnarray}
That is, our analysis of the MJ matrix suggests that  on average four 
times as many hydrophobic residues are buried in the core than  
are polar residues.  

\medskip
\noindent{\bf Two-body terms}.  
We define the true two-body part of the MJ matrix to be the matrix
minus the one-body and constant part of Eq.~(\ref{li_eq}): $M_{ij} -
c_0 - c_1(q_i + q_j)$.  This two-body part is again well approximated
by $c_2' q_i q_j$, $c_2' = -10.7$, with which it has a linear
correlation of 0.832.  When $c_2' q_i q_j$ is re-expressed in terms of
$Q^*$ using Eq.~(\ref{xi_eq}) the shift $q_0$ induces an additional
one-body term such that
\begin{equation}
M_{ij} \cong C_2~ \xi_i \xi_j Q^*_i Q^*_j + 
c'_1 (q_i + q_j) + {\rm const.}
\label{mjQq_eq}
\end{equation}
Where $C_2 = c'_2/c_1^2 = -0.423$ and $c_1' = c_1 - c'_2 q_0 = 5.62$.
The linear correlation between $M_{ij} - c'_1 (q_i + q_j)$ and $\xi_i
\xi_j Q^*_i Q^*_j$ is 0.681, see Fig.~1(b).  Given that the dipole
moments and $\xi_h$ and $\xi_p$ are predetermined, the first term on
the right-hand-side of Eq.~(\ref{mjQq_eq}) is a {\it one} free
parameter ($C_2$) fit to 210 pieces of ``noise'' in the MJ matrix.
The mediocre quality of the correlation nevertheless suggests that the
two-body term cannot be explained by dipole interactions alone;
interactions depending on charge and polarizability may need to be
included.  The inclusion of such terms may cause the two-body term 
to deviate from having the simple $qq$ form suggested by in Eq.~\ref{li_eq}.   
Owing to its relative small magnitude such a deviation 
should be tolerable to the original MJ matrix. 

\medskip
\noindent{\bf MJ matrix in terms of $Q^*$}.  
Re-expressing the one-body term in Eq.~(\ref{mjQq_eq})  
in terms of $Q^*$ and rationalizing notations by writing $\mu_{ij} =
C_2 \xi_i \xi_j$ and $\xi'_i = \xi_i c'_1/c_1$ 
we finally have  
\begin{equation} 
M_{ij} \cong \mu_{ij} Q^*_i Q^*_j    
+ \xi'_j Q^*_j + \xi'_i Q^*_i + \hbox{\rm const.} 
\label{final_eq}
\end{equation}
where $\mu_{hh} = -0.13$, $\mu_{hp} = -0.032$, $\mu_{pp} = -0.0078$,
$\xi'_h = 0.63$ and $\xi'_p = 0.15$.  The two sides of the equation
has a linear correlation of 0.922, see Fig.~1(c).  Since $Q_i$ is
either zero or positive, the negative values of $\mu_{ij}$ imply that
the dipoles mostly succeed in causing the residues to
lower their energies.  That is, even in a folded state the
the residues appear to be sufficiently unrestricted to find optimum
orientations.  To the extent that the dipole moments of the
side-chains are not free parameters, the expression on the
right-hand-side is a {\it four} parameter fit - $C_2$, $E_0$, $\Delta
N_h/ N_h$ and $\mu_w$ (see below) - to the complete MJ matrix.

\medskip
\noindent{\bf Residue-residue dipole coupling}.  
By  definition $\mu_{ij} \propto (\Delta N_i/
N_i)(\Delta N_j/ N_j)$.   With  $\Delta N_i/ N_i$ describing the 
percentage of buried  residues in a folded protein, the inequalities 
$|\mu_{pp}| < |\mu_{hp}| < |\mu_{hh}| < |\mu_r|$ correctly take into 
account the dielectric property of water: 
the coupling between residues shielded from water 
is stronger than that between residues that are not.   
The magnitude of the weighed average of the residue-residue coupling,  
$\bar \mu_{ij} = (7 \mu_{pp} +6 \mu_{hp} +7 \mu_{hh})/20
= -0.041$, is about four times less than the bare 
coupling strength of $|\mu_r| = 0.172$.  

\medskip
\noindent{\bf Water-residue coupling}.  We can obtain the effective 
water-residue coupling from the relation $\xi_i = -\mu_w (\Delta
N_i/N_i) (N_w Q_w/N_c)$ given earlier.  Using the value 6.5~$\AA$ for
the average effective diameter of a residue and the value 2~$\AA$ for
the diameter of a water molecule, we estimate that a residue may have
a maximum of 12 residue contacts and 57 water molecule contacts.  In
practice the number of contacts is encumbered by the presence of the
peptide backbone and geometric constraints, such that in fact 
$N_c \approx 7 \cite{Miyazawa85}$.  We therefore scale $N_w$ 
down to $\approx 35$.
With $Q_w = 1.85~D$, we deduce from Eqs.~(\ref{hidden_eq}) and
(\ref{final_eq}) that $\mu_w \approx -0.076\ (RT)$.  The negative sign
of $\mu_w$ is consistent with the notion that the presence of dipole
in a residue reduces its hydrophobicity.  Taking the average
water-residue distance to be 4.25~$\AA$ we expect the bare
water-residue coupling to be $(6.5/4.25)^3 = 3.5$ times stronger than
the bare residue-residue coupling.  However, in an unfolded state the
residues are completely exposed to water.  We therefore expect 
the approximate relations 
$|\mu_{pp}| < |\mu_w|/3.5 \cong |\mu_{hp}| \cong |\bar \mu_{ij}| <
|\mu_r|$, which are satisfied. 

\medskip 
\noindent{\bf Solvation energy, surface tension and hydrophobicity}.  
With $\mu_w$ and $Q_0$ extracted from the data we now find the bare
solvation energy to be $E_0 = \mu_w Q_0 Q_w N_w = 14.6\ RT$.  Although
hydration is an exceedingly complex process and is not fully
understood, the effective surface tension of water, or surface free
energy cost to water forced to sit against a hydrophobic surface has
been estimated to be $\sigma = 40$ erg/cm$^2$ \cite{Wortis}.  For a
residue of diameter $R_0$ the free energy cost is $W = 4 \pi (R_0/2)^2
\sigma = 13\ RT$, which is reasonably close to the value of $E_0$.
The fact that a good fit to the MJ matrix demands that 
$E_0$ enters $\Delta U$ in Eq.~\ref{energy2_eq} multiplied by 
$\Delta N_i$ is indication that $E_0$ needs to be surface energy. 
When the water-residue dipole interaction energy is included, the
total solvation energies $E_i$ of the residues then delineate into
groups with distinct hydrophobicities, with the seven most hydrophobic
(hydrophilic) having an average solvation energy of $13.2\ RT$ ($-
9.3\ RT$).

\medskip 
Very recently Keskin {\it et al.} \cite{Keskin} re-analyzed the MJ
matrix and derived the approximation (for ease of discussion the
$W^*_i$ used here has an additional negative sign relative to that in
\cite{Keskin}): 
$M_{ij} \cong \Delta W^*_{ij} + W^*_i + W^*_j + {\rm const.}$, 
where the one-body term $W^*$ is essentially defined as the
mean-field of $M_{ij}$ and $\Delta W^*_{ij}$ is a four parameter fit
to $M_{ij}$ minus its mean-field.  The analysis confirms the dominance
of the one-body term in the MJ matrix.  The overall fit to the MJ
matrix, with a correlation of 0.99, is excellent and the fit to the
two-body part is about the same as that given by the dipole picture:
the correlation between $M_{ij} - W^*_i - W^*_j$ and $\Delta W^*_{ij}$
is 0.67.  Not surprisingly $W^*$ and $q$ are closely related.  The
expression $\eta c_1 q + 1.16$, with scale factor $\eta=1.17$,
reproduces $W^*$ with a linear correlation of 0.997.  The value of
$\eta$ is mostly explained by the fact that the mean-field calculated
from the right-hand-side of Eq.~(\ref{li_eq}) is 
$1.22 c_1 (q_i + q_j)$.  
Incidentally, $\eta c_1 = 5.89$ is very close to the value of
the renormalized coefficient $c'_1 = 5.62$ given in
Eq.~(\ref{mjQq_eq}).

In Table I are listed values for $q$, $Q$, $W^*$, $\xi Q^*$, and
hydropathy scales $\Delta G$ (in units of $RT$) corrected for
self-solvation for the side-chains of the twenty amino acids
\cite{Roseman}.  Recall that $\xi$ contains the burial factor (see
Eq.~\ref{xi_ex_eq}) and $Q^*$ is $Q$ shifted by an amount proportional
to $E_0$ (see Eq.~\ref{xi_eq}).  The pairwise linear correlation of
the entries in Table I are given in column 2 of Table II. The
correlation between $\xi Q^*$ and $W^*$ (and $q$) is very
significantly better than that between $Q$ and $W^*$ (and $q$).  The
linear relations connecting the solvation energy with $\xi Q^*$, $W^*$
and $q$: $E_i (\Delta N_i/ N_i)/N_c = \xi_i Q^*_i \cong c_1 (q_i-q_0)
\cong (W^*_i -W^*_0)/\eta$, where $W^*_0 = 0.71$ is a shift, highlight
the importance of taking into account the burial factor of a residue
in a folded protein when interpreting the one-body terms of the MJ
matrix.

The hydropathy scales shown in Table I are derived for side-chains in
model peptides rather than in proteins.  They include the effect of
self-solvation that reduces the hydropathies of the polar
side-chains \cite{Roseman}, but does not include the effect of burial
factor.  This probably explains why, as seen in Table II, the $\Delta
G-q$, $\Delta G-W^*$, $\Delta G-Q$ and $\Delta G-\xi Q^*$ correlations
are of similar quality.

The $q$ and $W^*$ values of proline suggest it to be polar, while its
$Q$, $\xi Q^*$ and $\Delta G$ values say it is ambivalent or even
hydrophobic.  The third column in Table II shows that the
correlations listed either remain unchanged or improve when proline is
excluded from the linear fit.  The ambiguous hydrophobicity 
of this residue may be related to the fact that is has a 
looping structure. 

We summarize our interpretation of Eq.~(\ref{li_eq}) being a good
approximation of the MJ matrix as follows.  The one-body part, or
hydrophobicity (or hydropathy) energy, is made up of two parts: free
energy cost to water to accommodate the residue surface, and
attractive dipole interaction between residue and water.  Because
polar residues have large dipole moments, hydrophobic residues have
small or no moments and ambivalent residues have something in between,
the hydropathic/hydrophobic energy is strongly attractive, weakly
attractive and strongly repulsive for polar, ambivalent and
hydrophobic residues, respectively.  Residue-residue dipole
interactions accounts for a sizable portion, but not all, of the
two-body part.  Aside from using the given dipole moments for the
residues and having two burial factors, one each for the hydrophobic
and polar residues, no residue-dependent adjustments were made in
deriving Eq.~(\ref{final_eq}), our rendition of Eq.~(\ref{li_eq}).
That is, we have not attempted a detailed fit of the MJ matrix.  The
correlation between the dipoles of the residues and $q$ becomes
unequivocal and the strengths of the dipole couplings extracted from
the MJ matrix become reasonable only when the burial factors are
included in the formulation.  That the factor is important reveals the
dynamical nature of protein folding: strengths of interactions change
as the folding progresses.  Protein folding is a very complicated
process that depends on many details and the MJ matrix does not tell
its whole story.  It does however contain the most basic structural
information at the molecular level of those proteins whose structures
are known.  The success of the present analysis in understanding the
main features of the MJ matrix gives us confidence that the model used
here may provide a starting point for building a true potential
suitable for use in a molecular dynamical description of early folding
of protein in water.

\medskip
The authors thank C. Tang, G.M. Crippen, Y. Duan, M. Wortis, D.C.Y. Lu and
P.G. Luan for discussions and the referee for pointing them to reference
\cite{Keskin} and for useful suggestions.  HCL thanks the Physics
Department of Simon Fraser University for hospitality in the Summer of
1998 during which part of the paper was written.  This work is partly
supported by grant NSC88-M-2112-008-009 from National Science Council
(ROC).


\begin{table}[hbt]
\caption{\noindent Values for $q$'s, $Q$'s (in Debye), 
 $W^*$, $\xi Q^*$, and 
$\Delta G$ (self-solvation corrected hydrophobicities); see text.}
\label{table1} 
\vspace{4pt}
\begin{tabular}{cccccc} 
Res.& $q$ & $Q$ & $W^*$ & $\xi Q^*$ & $\Delta G$\\
\hline
 Cys & -0.265 &  0.540 & -0.246 & -1.36 & -3.33\\ 
 Met & -0.327 &  0.218 & -0.707 & -1.54 & -2.78\\
 Phe & -0.438 &  0.393 & -1.512 &  -1.44 &  -5.40\\
 Ile & -0.390 &  0.046 & -1.087 &  -1.63 &  -5.03\\
 Leu & -0.443 &  0.006 & -1.502 &  -1.66 &  -5.03\\
 Val & -0.315 &  0.021 & -0.633 &  -1.65 &  -3.63\\
 Trp & -0.298 &  0.762 & -0.656 &  -1.23 &  -4.77\\
 Tyr & -0.226 &  2.40  & -0.355 &  -0.315  &   1.63\\
 Ala & -0.125 &  0.00  & 0.531 &  -0.403 &  -1.12\\
 Gly & -0.048 &  0.00  & 0.845 &  -0.403 &  0.00\\
 Thr & -0.058 &  2.39  & 0.828 &  -0.078 &  -0.70\\
 Ser & -0.011 &  2.40  & 1.076 &  -0.076  &  0.17\\
 Asn & -0.011 &  4.03  & 1.104 &  0.145  &  3.78\\
 Gln & -0.023 &  3.81  & 1.038 &  0.116  &  3.53\\
 Asp &  0.040 &  4.29  & 1.302 &  0.180  &  2.62\\
 Glu &  0.028 &  6.08  & 1.334 &  0.424  &  2.97\\
 His & -0.107 &  2.85  & 0.429 &  -0.014  &  1.82\\
 Arg & -0.020 &  4.90  & 1.043 &  0.264  &  6.48\\
 Lys &  0.065 &  8.09  & 1.648 &  0.697  &  4.10\\
 Pro & -0.054 &  1.40  & 0.907 &  -0.212  &  -2.92 
\end{tabular}
\end{table}

\begin{table}[hbt]
\caption{Linear correlations.} 
\label{table2} 
\vspace{4pt}
\begin{tabular}{lcc} 
Pair entries & Correlation & Correlation w/o Pro\\
\hline
$\phantom{W^*}q$ $vs.$ $W^*$  & 0.997 & 0.997 \\
\hline
$\phantom{W^*}q$ $vs.$ $Q$ & 0.753 & 0.775 \\
$\phantom{q}W^*$ $vs.$ $Q$ & 0.743 & 0.767 \\
$\phantom{W^*}q$ $vs.$ $\Delta G$ & 0.836 & 0.880 \\
$\phantom{q}W^*$ $vs.$ $\Delta G$ & 0.820 & 0.866 \\
$\phantom{W^*}Q$ $vs.$ $\Delta G$ & 0.843 & 0.839 \\
\hline
$\phantom{W^*}q$ $vs.$ $\xi Q^*$ & 0.949 & 0.949 \\
$\phantom{q}W^*$ $vs.$ $\xi Q^*$ & 0.932 & 0.933 \\
$\Delta G$ $vs.$ $\xi Q^*$ & 0.890 & 0.923 
\end{tabular}
\end{table}

\end{document}